\documentclass{aa}  

\usepackage{graphicx}
\usepackage{txfonts}
\begin{document}

   \title{Chemical abundance analysis of the old, rich open cluster  Trumpler~20\thanks{Based on observations collected at Paranal
       Observatory under program 088.D-0045}}

\author{Giovanni Carraro\thanks{On leave from the Dipartimento di Fisica e Astronomia, Universit\`a di Padova, Italy} 
          \inst{1}, Sandro Villanova\inst{2}, Lorenzo Monaco\inst{1},\\ Giacomo Beccari\inst{1},
          Javier A. Ahumada\inst{3}, and Henri M.J. Boffin\inst{1}
          }

     \institute{ESO, Alonso de Cordova 3107, 19001,
           Santiago de Chile, Chile\\
              \email{gcarraro,lmonaco,gbeccari,hboffin@eso.org}
         \and
            Departamento de Astronomia, Universidad de Concepci\'on,
               Casilla 169, Concepci\'on, Chile\\
               \email{svillanova@astro-udec.cl}
             \and
           Observatorio Astron\'omico, Universidad Nacional de C\'ordoba,
           Laprida 854, 5000
               C\'ordoba,  Argentina\\
             \email{javier@oac.uncor.edu}
             }
\authorrunning{Carraro et al.}

   \date{Received October 1, 2013; accepted October 16, 2013}

 
  \abstract
   {}
   {Trumpler~20 is an open cluster located at low Galactic longitude, just beyond the great Carina spiral arm, and whose metallicity and fundamental parameters were very poorly known until now. As it is most likely a rare example of an old, rich open cluster -- possibly a twin of  NGC 7789 -- it is useful to characterize it. To this end, we
   determine here the abundance of several elements and their ratios in a sample of stars in the clump of Trumpler~20. 
   }
   {We present high-resolution spectroscopy of  eight clump stars. Based on their radial velocities, we identify six {\it bona fide} cluster members, and for
  five of them (the sixth  being a fast rotator) we perform a detailed abundance analysis.}
   { We find that Trumpler 20 is slightly more metal-rich than the Sun, having  [Fe/H]=+0.09$\pm$0.10. The abundance ratios of $\alpha$-elements are generally solar. In line with recent studies of clusters as old as Trumpler~20, Ba is overabundant compared to the Sun. Our  analysis of the iron-peak elements 
   (Cr and  Ni) does not reveal anything anomalous. Based on these results, we re-estimate the cluster age to be 1.5$^{+0.2}_{-0.1}$ Gyr. Its distance to the Galactic centre turns out to be 7.3 kpc. With this distance and metallicity, Trumpler~20 fits fairly well in the metallicity gradient for the galactic inner disc. }
   {With this new study, the characterization of Trumpler~20 is now on  much more solid ground. Further studies should focus on the estimate of the binary fraction and on its main sequence membership.}


   \maketitle
%

\section{Introduction}
High-resolution spectroscopic data of stars in Galactic old open clusters are accumulating
very rapidly. These star clusters  are widely recognized as ideal tracers of the Galactic disc chemical and dynamical
evolutions. A precise knowledge of the star cluster's abundance of the various chemical elements and their ratios  allows us to investigate
the assembly history of the thin disc,  its relationship with the thick disc and the bulge, and the presence of sub-populations,
which may be indicative of past accretion events (Friel 1995, Carraro et al. 2007, Magrini et al. 2009).\\

Rich old open clusters are particularly interesting, since the most important evolutionary phases
are better represented in the colour-magnitude diagram, and allow better comparison with theoretical models
of stellar evolution.
Rich clusters  are not very common among old open clusters, since the typical lifetime of a star cluster is around 200 Myr (Gieles et al. 2011).
Typical examples are NGC 7789 (Gim et al. 1998) and NGC 2158 (Carraro et al. 2002).

A new member of this small family has recently been  identified: Trumpler ~20 (Platais et al. 2008, Seleznev et al. 2010a).
This cluster possesses a conspicuous clump of He-burning stars, as do NGC 2158 and NGC 7789, but its main sequence (MS)
is severely blurred by interlopers from the Galactic disc, and by some differential reddening (Platais et al. 2012). 
This is mostly because of its location
just beyond the great Carina spiral arm, and its relatively low Galactic latitude ($l = 301.5^\circ, b = 2.2^\circ$).
This has prevented a precise estimate of its age, since the location of the MS turnoff point (TO) is difficult to detect.
Binary stars also play a crucial role (Carraro et al. 2010). 
The interest in Trumpler~20 lies mostly in its clump, as stressed by Carraro et al. (2010). Depending on its age,
Trumpler~20 might actually  be a twin of NGC~7789, and would  therefore help to understand the details of the He-burning phases
in the corresponding mass range (Girardi et al. 2000).

As described above, it has been quite difficult to estimate the cluster's parameters, especially its age.
Platais et al. (2008) and Seleznev et al. (2010) first recognized the potential interest of Trumpler~20.
Platais et al. (2008) concluded that the cluster is 1.3 Gyr old, metal poor ([Fe/H]  $\sim -0.11$ dex), and at a distance
of 3.3 kpc, adopting a reddening of $E(B-V) = 0.46$.
Seleznev et al. (2010)  confirmed this set of parameters, ,within the {\it large} uncertainties,  by assuming solar abundances
and $V, I$ photometry. Later on, Carraro et al. (2010) provided a more detailed study of Trumpler 20, based on $UBVI$ photometry.
This study, which also assumed solar metallicity,  revised the cluster parameters: the reddening value was found to be significantly smaller, $E(B-V) =0.35$, 
the distance around 3 kpc, and the age around 1.4 Gyr. 

\begin{figure}
\includegraphics[width=\columnwidth]{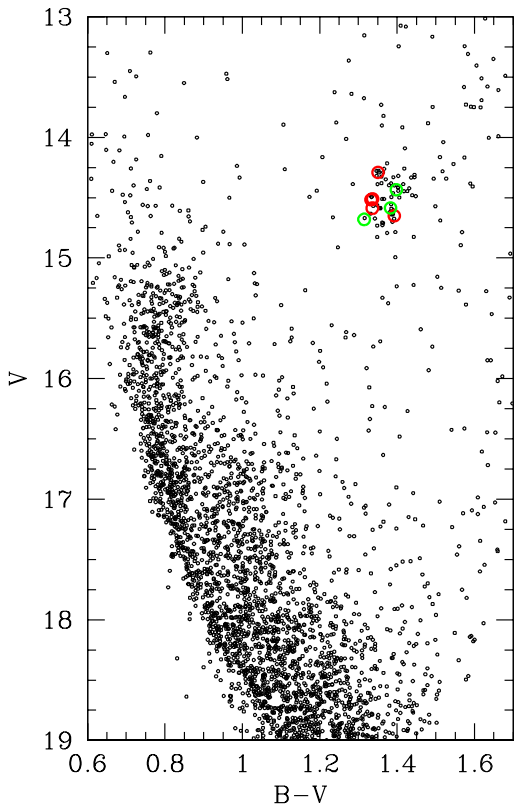}
\caption{$V~ vs.\,B-V$ colour-magnitude diagram of Trumpler~20 from Carraro et al. 2010.
Red symbols indicate {\it bona fide} members.}
\end{figure}

Clearly, the lack of a precise metallicity estimate from high-resolution spectra limited the study of this cluster. 
The recent study by Platais et al. (2012) did not
solve this problem. The study concentrated on the upper MS of the cluster in an attempt to clean the TO region by using
medium-resolution spectra. The upper MS, once corrected for membership and differential reddening, is then fitted
by using a Z =0.015 isochrone, which implies the same set of parameters as in Platais et al. (2008), except for the reddening,
which is closer to the Carraro et al. (2010) estimate. 

In an attempt to improve the cluster characterization, in this paper we present the result of a high-resolution spectroscopic
observational campaign aimed at determining the abundance of many elements for the stars in the red clump of Trumpler~20. 
The paper is organized as follows.  In Section~2 we present the observational material and describe how radial velocities are derived.
Section~3 is devoted to the abundance analysis, while in Section~4 we discuss in detail the results of the analysis.
Section~5 summarizes our findings.


\begin{table*}[hbpt]
\caption{Red clump stars observed with UVES in the field of Trumpler 20. The two epoch radial velocities (RV) are listed together with their mean values.}
\begin{tabular}{cccccrrr}
\hline \hline
  ID   &       RA &      DEC &      $V$ &   $B-V$      &  RV~~~~~~~~& RV~~~~~~~~ &   <RV>~~~~ \\
  & (J2000.0) & (J2000.0) & & & (2012/02/12)  & (2012/03/04) & \\
  \hline
          &  hh:mm:ss.s  &  dd:mm:ss.s   &      mag           &   mag   &   (km\,s$^{-1}$)~~~ & (km\,s$^{-1}$)~~~ & (km\,s$^{-1}$)~~~\\      
\hline              
  80  &  12:40:11.4   &  $-$60:30:42.8   &	     14.73   &   1.40   &    $-$38.01$\pm$1.17  & $-$37.10$\pm$1.04  & $-$37.56$\pm$0.57 \\
 207  &  12:39:55.6   &  $-$60:37:26.8   &	     14.70   &   1.33   &    $-$39.69$\pm$1.30  & $-$39.89$\pm$1.15  & $-$39.79$\pm$0.13 \\   
 258  &  12:39:17.5   &  $-$60:33:35.4   &	     14.47   &   1.35   &    $-$41.13$\pm$1.45  & $-$41.16$\pm$1.23  & $-$41.14$\pm$0.02  \\
 429  &  12:40:01.2   &  $-$60:32:49.2   &           14.61   &   1.39   &       57.40$\pm$2.22  &    75.15$\pm$2.15  &                     \\
 496  &  12:39:37.7   &  $-$60:36:02.2   &	     14.86   &   1.31   &    $-$46.67$\pm$4.57  & $-$44.81$\pm$4.02  & $-$45.74$\pm$1.17   \\
 542  &  12:39:12.0   &  $-$60:36:32.2   &	     14.69   &   1.33   &    $-$40.85$\pm$1.25  & $-$40.94$\pm$1.12  & $-$40.90$\pm$0.06   \\
 609  &  12:39:40.0   &  $-$60:37:09.6   &	     14.77   &   1.33   &    $-$40.38$\pm$1.22  & $-$40.64$\pm$1.12  & $-$40.52$\pm$0.17   \\
5309  &  12:40:44.7   &  $-$60:37:32.0   &           14.77   &   1.37   &       11.53$\pm$1.21  &    11.07$\pm$1.17  &    11.30$\pm$0.29   \\
\hline\hline
\end{tabular}
\end{table*}

\section{Observation and data reduction}

Observations were taken in service mode on the nights of February 11 and March
3, 2012 using the multi-object fibre-fed FLAMES facility mounted at the
ESO-VLT/UT2 telescope at the Paranal Observatory (Chile).  Two 2400s exposures
were taken simultaneously with the  GIRAFFE medium-resolution spectrograph, and
the red arm of the UVES high-resolution spectrograph. GIRAFFE was configured in
Medusa mode in the setup HR08, which covers the wavelength range 
4911--5163\AA\, at a resolution of $R$=20,000. The UVES spectrograph was, instead,
set up around a 5800\AA\, central wavelength, thus covering the 4760--6840\AA\, wavelength
range and providing a resolution of R$\simeq$47,000.

Here we discuss UVES data.UVES targets observed the  probable
red-clump targets, and are shown as open circles at the top of the cluster's  $V\, vs.~ B-V$
colour-magnitude diagram in Fig.~1.  In this figure red symbols indicate  the five radial velocity members,
and green the three stars for which abundance analysis was not done  because they are non-members or
fast rotator/binaries (see text for more details). 
Table\,1  presents the target stars' IDs,
coordinates, and $B$ and $V$ photometry from  Carraro et al. (2010).
The data were reduced using the ESO CPL based FLAMES-UVES pipeline version
5.0.9\footnote{\url{http://www.eso.org/sci/software/pipelines/}} for
extracting the individual fiber spectra. 

The spectra were eventually normalized using the standard IRAF task {\tt
continuum}. Radial velocities were computed using the IRAF/{\tt fxcor} task to
cross-correlate the observed spectra with a synthetic one from the
\citet[][]{coelho05} library with stellar parameters  $T_{\rm eff}$=5250\,K, log\,$g$=2.5, solar
metallicity, and no $\alpha$-enhancement. The IRAF {\tt rvcorrect} task was used
to calculate the correction from geocentric velocities to heliocentric. We took  the star's radial
velocity to be the average of the two epochs measured and the error to be the difference
between  the two values multiplied by 0.63 \citep[small sample
statistics; see ][]{keeping62}. For star 429 we report both measured values together
with the formal {\it fxcor} uncertainties. The velocity difference between the two
measurements (17.75\,\,km\,s$^{-1}$) is significantly higher than the estimated
errors and, therefore, the star may be in a binary system. Stars 429 and
496 are probably rotating faster than the other stars. The FWHM of the
cross-correlation function is about 21\,\,km\,s$^{-1}$ and 33\,\,km\,s$^{-1}$
for the two stars, respectively, to be compared to an average for the other
stars of 14.2$\pm$0.2\,\,km\,s$^{-1}$. \\
\noindent
We exclude stars 429 and 5309,
which are non-radial velocity members, and using the non-parametric jackknife resampling method (Lupton 1993)
we obtain for the cluster an
average radial velocity of $<V_r>=-40.9\pm$1.2\,km\,s$^{-1}$ with a dispersion 
 $\sigma$=2.7$\pm$1.4\,km\,s$^{-1}$. \\
Had we also excluded star 496, which is a fast rotator,  we would have obtained
$< Vr >=-40.0\pm 0.7$\,km\,s$^{-1}$, and  $\sigma=1.4\pm0.9$\,km\,s$^{-1}$ .

\noindent 
These values is in good
agreement with that measured by Platais et al (2008;2012), i.e.  $-40.8$ and $-40.6$\,km\,s$^{-1}$,
respectively.

Finally, for the abundance analysis, the two epoch rest-frame spectra obtained for each star, were averaged together.
The final spectra have signal-to-noise (SNR) ratios in the range 30-50 at $\sim$6070\AA.

 \begin{table*}
\caption{Atmospheric parameters from photometry (ph) and spectroscopy (sp).}
\begin{tabular}{cccccccc}
\hline \hline
  ID   &       $T_{\rm eff, ph}$ &      $\log g_{\rm ph}$ &      $T_{\rm eff,sp}$ &   $\log g_{\rm sp}$     &  $v_t$ &  [Fe/H] & Comment\\
  \hline
          &   K  &    &     K     &     &   (km\,s$^{-1}$) & dex\\      
\hline              
  80   &  4716   &  2.60  &	 4970     &   2.55   &   1.11 &   0.09&\\
 207  &  4847   &  2.73  &	 4920     &   2.52   &   1.14 &   0.08&\\  
 258  &  4812   &  2.86  &	 4910     &   2.50   &   1.18 &   0.06&\\
 429  &  4730   &  2.65  &   4670     &   2.59   &   1.67 &  -0.41& non-member\\
 496  &  4876   &  2.63  &	               &              &            &          & fast rotator \\
 542  &  4840   &  2.76  &	 4950     &   2.55   &   1.13 &  0.11&\\
 609  &  4840   &  2.75  &	 4970     &   2.60   &   1.10 &  0.10&\\
5309 &  4762   &  2.84  &   4650     &   2.75   &   0.80 &  0.08& non-member\\
\hline\hline
\end{tabular}
\end{table*}

\begin{table*}
\caption{Individual abundances for Trumpler 20 members. Apart from iron, abundances are provided as [X/Fe], where X is the chemical species.}
\begin{tabular}{crrrrrrrrrr}
\hline\hline
ID    & C     & N   &  O &   CNO  &  Na  & Mg &  Al &   Si &   Ca  &  Ti\\
\hline
  80& $-$0.20&  0.04&     $-$0.20& $-$0.17& 0.20& $-$0.11& 0.01&  0.02     & $-$0.06& $-$0.10\\
207& $-$0.18&  0.04&     $-$0.15& $-$0.14& 0.22& $-$0.16& 0.01& $-$0.01& $-$0.02& $-$0.06\\
258& $-$0.10& $-$0.08& $-$0.10& $-$0.10& 0.20&      0.00& 0.08&  0.02     &    0.02& $-$0.01\\
542& $-$0.21&  0.00&      $-$0.23& $-$0.20& 0.21& $-$0.15& 0.10&  0.00   &  0.04& $-$0.01\\
609& $-$0.22&  0.00&     $-$0.23& $-$0.20& 0.20& $-$0.09& 0.08&  0.00    &  0.00&  \\
\hline\hline
ID    & V &  Cr &  [Fe/H] &  Ni &   Y &   Zr &  Ba &   La &   Eu&\\
\hline
  80& 0.45& 0.08& 0.08& 0.05& $-$0.16& 0.10& 0.19&     & $-$0.03&\\
207& 0.56& 0.00& 0.09& 0.05&  $-$0.17& 0.05& 0.19& $-$0.05&  0.00&\\
258& 0.47& 0.02& 0.06& 0.02&           & 0.05& 0.24& $-$0.04& $-$0.03&\\
542& 0.37& 0.03& 0.11& 0.06&  $-$0.09& 0.04& 0.13& $-$0.03& $-$0.07&\\
609& 0.39& 0.05& 0.10& 0.02&  $-$0.14& 0.07& 0.17& $-$0.01& $-$0.08&\\
\hline
\end{tabular}
\end{table*}

\begin{table*}
\caption{Number of lines, and derived abundance $\sigma$. The value 9.999 indicates that $\sigma$ was not computed.}
\begin{tabular}{crrrrrrrrrrrrr}
\hline\hline
ID  &   FeI &    FeII&    Na &    Mg &    Al &    Si &    Ca &    TiI&     TiII&    V &     Cr &     Ni&     Y\\
\hline
207& 120/0.12&   9/0.12&  2/0.05& 2/0.19& 1/9.99& 9/0.08& 6/0.15& 15/0.18& 3/0.15& 1/9.99& 13/0.23& 29/0.12& 3/0.13\\
258& 123/0.13& 11/0.16&  2/0.08& 1/9.99& 2/0.21& 7/0.11& 6/0.23& 11/0.14& 3/0.04& 1/9.99& 11/0.21& 28/0.14& 0/9.99\\
542& 130/0.12& 12/0.08&   2/0.04& 1/9.99& 2/0.06& 8/0.09& 7/0.18& 13/0.16& 3/0.20& 1/9.99& 11/0.13& 30/0.16& 3/0.03\\
609& 122/0.13&   9/0.13&  2/0.08& 1/9.99& 2/0.19& 8/0.12& 7/0.19& 12/0.17& 3/0.22& 1/9.99& 8/0.12& 27/0.14 & 3/0.15\\
80  & 122/0.14&    9/0.13&  2/0.15& 1/9.99& 2/0.15& 8/0.09& 6/0.11& 11/0.18& 2/0.02& 1/9.99& 8/0.28& 26/0.17&  3/0.24\\
\hline\hline
ID    &  C     &  N    &  O    &  Ba  &   Eu    & La &   Zr\\
\hline
207& C2-band &CN-band &1/9.99 &1/9.99 &1/9.99 &1/9.99 &1/9.99\\
258& C2-band &CN-band &1/9.99 &1/9.99 &1/9.99 &1/9.99 &1/9.99  \\
542& C2-band &CN-band &1/9.99 &1/9.99 &1/9.99 &1/9.99 &1/9.99\\
609& C2-band &CN-band &1/9.99 &1/9.99 &1/9.99 &1/9.99 &1/9.99\\
80&  C2-band &CN-band &1/9.99 &1/9.99 &1/9.99 &1/9.99 &1/9.99  \\
\hline
\end{tabular}
\end{table*}

\section{Abundance analysis}

The chemical abundances for Na, Mg, Al, Si, Ca, Ti, Cr, Fe, and Ni were
obtained using the equivalent widths (EW  method, as detailed in 
 \citet{Ma08}.
For C, N, O, Y, Ba, La, and Eu, whose lines are affected by blending, we used the
spectrum-synthesis method. For this
purpose we calculated five synthetic spectra having different abundances for the
elements, and estimated the best-fitting value as the one that minimizes the
$r.m.s.$ scatter. Only lines not contaminated by telluric lines were used.
ATLAS9 (Kurucz 1970) model atmospheres were used for the methods:
EW and spectrum-synthesis. 
The initial atmospheric parameters for the model atmosphere 
were assumed to be  those typical for a RGB star of an open cluster, i.e.
$T_{\rm eff}$=4500 \,K, log\,$g$=2.5, $v_{\rm t}$=1.20 km/s, and [Fe/H]=0.0.
We then  refined them during the abundance analysis. As a first step, atmospheric models were calculated using ATLAS9 \citep{Ku70}
using the initial estimates of $T_{\rm eff}, \log g$,
 $v_{\rm t}$, and  [Fe/H].
 
The value of T$_{\rm eff}$, v$_{\rm t}$, and $\log g$ were adjusted and new
atmospheric models calculated in an interactive way in order to remove trends in
excitation potential (EP) and equivalent width $vs.$ abundance for
T$_{\rm eff}$ and v$_{\rm t}$, respectively, and to satisfy the ionization
equilibrium for $\log g$. \ion{Fe}{I} and \ion{Fe}{II}  were used for this purpose.
The [Fe/H] value of the model was changed at each iteration according to the
output of the abundance analysis.
The local thermodynamic equilibrium (LTE) program MOOG \citep{Sn73} was used
for the abundance analysis.

Typical internal errors are $\Delta(T_{\rm eff})$=50 K,  $\Delta\log g$=0.2 dex, $\Delta v_{\rm t}$=0.10 km/s, 
and $\Delta$[Fe/H]=0.05 dex, respectively. 
To better clarify the effect of atmospheric parameters on the derived abundance, we consider star 542, and show the effect
on the final abundance of changes in the temperature, gravity, and micro-turbulence velocity. The results are shown in
Table~5.

\begin{table}
\caption{Error budget analysis. Under $El$ the ratio $\Delta$ [X/Fe] is indicated, except for Fe, where $\Delta$[Fe/H] is shown.}
\begin{tabular}{cccc}
\hline\hline    
 El.&      $\Delta$ T=+50K  & $\Delta$ log(g)=+0.20 &  $\Delta$ vt=+0.10\\
\hline  
Fe    &  +0.03 &     +0.00    &    -0.07\\
FeII  & -0.10     &  +0.09    & -0.05\\
Na   &  +0.01  &    -0.03     &   +0.03\\
Mg   &  -0.01   &   -0.07    &   +0.00\\
Al     & +0.01  &    -0.01    &    +0.05\\
Si    & -0.05     &  0.02      &  +0.02\\
Ca  &  +0.07  &     0.03     &   +0.11\\
TiI   & +0.04   &   -0.01    &    +0.03\\
TiII  & -0.07    &   0.06     &   -0.04\\
V     & +0.05   &    0.01   &     +0.02\\
Cr   &  +0.03  &    -0.01  &      +0.03\\
Ni    & -0.01    &   0.03   &     +0.00\\
Y     & -0.06    &   0.06    &    -0.04\\
C     & -0.06   &    0.03    &    +0.04\\
N     & +0.00  &     0.03   &     +0.06\\
O     & -0.02    &   0.11   &     +0.04\\
Ba   &  -0.01   &    0.06  &      -0.03\\
Eu   &  -0.04   &    0.11  &      +0.05\\
La   &  -0.04   &    0.10  &      -0.01\\
Zr    & -0.05    &   0.11  &      +0.0\\
\hline
\end{tabular}
\end{table}

The line lists for the chemical analysis were obtained from many sources 
(Gratton et al. 2003, VALD \& NIST\footnote{\url{http://physics.nist.gov/PhysRefData/ASD/}}; 
McWilliam \& Rich 1994; McWilliam 1998, SPECTRUM\footnote{\url{http://www.phys.appstate.edu/spectrum/spectrum.html}}, 
and SCAN\footnote{\url{http://www.astro.ku.dk/~uffegj/}}), 
and the log\,gf were calibrated using the solar-inverse technique and by the spectral synthesis method (see Villanova et al. 2009  for more details). 
For this purpose we used the high resolution, high S/N NOAO solar spectrum (Kurucz et al. 1984). 
The solar abundances we obtained with our line list are reported in Table~4,
together with those given by Grevesse \& Sauval (1998) for comparison. 
We emphasize the fact that all the line-lists were calibrated on the Sun, including those used for the spectral synthesis.
 We provide it as a long table at the bottom of the paper.
In addition, the  C content was obtained from the C$_2$ system at 563.2 nm, and N from the CN lines
at 634 nm.\\

\noindent
Abundances for C, N, and O were determined all together in an
interactive way in order to take into account any possible molecular coupling
of these three elements.
Our targets are objects evolved off the main sequence, so some evolutionary
mixing is expected. This can affect the primordial C and N abundances separately,
but not the total C+N+O content because these elements are transformed one
into the other during the CNO cycle.\\

\noindent
We are aware that we are deriving abundances for clump
stars, but using the Sun as calibrator. Different approaches 
are adopted in the literature. Among the references we will use 
for comparison in the following, a similar approach was followed 
by Magrini et al. 2010, but not by Bensby et al. 2010.
As a sanity check, we computed  abundances for the giant star Arcturus using our line-list. Its atmospheric parameters
are  $T_eff=4290 ^oK$, logg=1.63,  and vt=1.29 km/sec.

The results for individual elements are shown in Table.~6.
One can readily see that the agreement with the literature 
is generally good. For instance, the differences with the recent 
work by McWilliam et al. 2013 are below 0.2\,dex. This is valid for 
all species except V, for which we obtain a significant difference that our value  is 
0.34\,dex higher than the one in McWilliam et al. 2013. A correction of this 
size would bring the V abundance of the cluster to a value more 
similar to thin disc stars (see Fig.~3). It should be noticed, however, 
that the Arcturus literature abundances are quite scanty for this element.
Other differences with McWilliam et al. 2013  are for Ti (0.13 dex), Y (0.1 7dex), and Ba (0.1 3dex).
Our [Y/Fe] value is right in-between the McWiliam et al. 2013 and Peterson et al. 1993 values.

\begin{table*}
\caption{Individual abundances for Arcturus. Columns 5 to 10 indicate literature values from McWilliam et al. (2013, McW13), 
Ramirez and Allende Prieto (2011, Ra11), Fulbright et al. (2007, Fu07), Worley et al. (2009, W09), and
Peterson et al. (1993, P93), respectively }
\begin{tabular}{cccccccccc}
\hline\hline
El     &  X & Sun  & [X/H] & [X/Fe] & McW13 &  Ra11  & Fu07  & W09  &   Pe93\\
\hline      		     		     			           			   
C       & 7.885$\pm$0.005& 8.49&  -0.60&  -0.10   &           &   +0.43&           &               &     +0.0\\
N       & 7.79& 7.95&  -0.16&  +0.34  &           &            &            &         	 &  +0.3\\ 
O       & 8.71& 8.80&  -0.09&  +0.41  & +0.46&  +0.40&  +0.48&  +0.57	  & +0.4\\
Na     & 5.92 & 6.32&  -0.26&  +0.10 &  +0.09&  +0.11&  +0.09&  +0.15	   &  \\
Mg     & 7.513$\pm$0.005&  7.56&  -0.05&  +0.45  & +0.39&  +0.37&  +0.39&  +0.34	  &    \\
Al       & 6.304$\pm$0.023& 6.48&  -0.18&  +0.32  & +0.38&  +0.34&  +0.38&  +0.25	   &   \\
Si       & 7.408$\pm$0.004& 7.61&  -0.21&  +0.29  & +0.35&  +0.33&  +0.35&  +0.20	  &    \\
Ca     & 6.101$\pm$0.017& 6.39&  -0.29&  +0.21  & +0.21&  +0.11&  +0.21&  +0.19	  &    \\
Ti       & 4.83& 4.94&  -0.11&  +0.39  & +0.26&  +0.27&  +0.26&  +0.35	  &    \\
V        & 4.00$\pm$0.01& 4.04&  -0.04&  +0.46  & +0.12&  +0.20&           &        	  &    \\
Cr      & 5.175$\pm$0.010& 5.63&  -0.45&  +0.05  &            &   -0.05&           &   	           &    \\
Fe      & 7.00$\pm$0.01& 7.50&  -0.50&              & -0.49  &  -0.52 &           & -0.61	  &    \\
Ni      & 5.786$\pm$0.008& 6.26&  -0.49&  +0.01  &            &  +0.06&           &   	            &    \\
Y       & 1.697$\pm$0.003& 2.25&  -0.55&  -0.05   & -0.22  &            &           &  +0.12     &  \\
Ba     & 1.795$\pm$0.005& 2.34&  -0.54&  -0.04   & -0.17  &            &           &  -0.19	   &   \\
Eu     & 0.32$\pm$0.05& 0.52&  -0.20&  +0.30  &+0.23  &            &           & +0.36	   &   \\
\hline
\end{tabular}
\end{table*}

\noindent

\section{Discussion}
In this section, we discuss in detail the outcome of the abundance analysis and its impact
on the determination of the cluster parameters.

\begin{table}
\caption{Adopted solar abundances}
\begin{tabular}{ccc}
\hline\hline
Element & This work & Grevesse \\
& & \& Sauval (1998)\\
\hline
Fe  &         7.50           &        7.50\\
C    &         8.49          &         8.52\\
N    &         7.95          &         7.92\\
O(6300)&   8.80      &             8.83\\
Na  &    6.32            &       6.33\\
Mg   &   7.56          &         7.58\\
Al    &  6.48           &        6.47\\
Si    &  7.61           &        7.55\\
Ca   &   6.39            &       6.36\\
Ti     & 4.94            &       5.02\\
V      & 4.04          &         4.00\\
Cr     & 5.63           &        5.67\\
Ni     & 6.26          &         6.25\\
Y      & 2.25           &        2.24\\
Zr     & 2.56           &        2.60\\
Ba    &  2.34       &            2.13\\
La     & 1.26      &             1.17\\
Eu     & 0.52    &               0.51\\
\hline
\end{tabular}
\end{table}

\begin{table}
\caption{Mean Trumpler 20 abundance ratios.}
\begin{tabular}{crr}
\hline\hline
Abundance ratio & Mean~~~ & $\sigma$~~~~~~ \\
\hline                
<[C/Fe]>  &         $-$0.18$\pm$0.02    & 0.05$\pm$0.02\\
<[N/Fe]>   &       0.00$\pm$0.02       & 0.05$\pm$0.02\\
<[O/Fe]>    &      $-$0.18$\pm$0.03      & 0.06$\pm$0.02\\
<[CNO/Fe]> &  $-$0.16$\pm$0.02  & 0.04$\pm$0.01\\
<[Na/Fe]>  &    0.21$\pm$0.00  & 0.01$\pm$0.00\\
<[Mg/Fe]>  &    $-$0.10$\pm$0.03  & 0.06$\pm$0.02\\
<[Al/Fe]>   &     0.06$\pm$0.02  & 0.04$\pm$0.01\\
<[Si/Fe]>   &     0.01$\pm$0.01  & 0.01$\pm$0.00\\
<[Ca/Fe]>  &     0.00$\pm$0.02  & 0.04$\pm$0.01\\
<[Ti/Fe]>    &       $-$0.03$\pm$0.02  & 0.04$\pm$0.01\\
<[$\alpha$/Fe]> &$-$0.01$\pm$0.01  & 0.02$\pm$0.01\\
<[V/Fe]>  &           0.45$\pm$0.03  & 0.07$\pm$0.02\\
<[Cr/Fe]>  &         0.04$\pm$0.01  & 0.03$\pm$0.01\\
<[Fe/H]>     &       0.09$\pm$0.01  & 0.02$\pm$0.01\\
<[Ni/Fe]>   &        0.04$\pm$0.01  & 0.02$\pm$0.01\\
<[Y/Fe]>    &        $-$0.14$\pm$0.02  & 0.04$\pm$0.04\\
<[Zr/Fe]>   &         0.06$\pm$0.01  & 0.02$\pm$0.01\\
<[Ba/Fe]>  &         0.18$\pm$0.02  & 0.04$\pm$0.01\\
<[La/Fe]>   &          $-$0.03$\pm$0.01  & 0.02$\pm$0.01\\
<[Eu/Fe]>   &        $-$0.04$\pm$0.01  & 0.03$\pm$0.01\\
\hline
\hline
\end{tabular}
\end{table}

\subsection{Metallicity}
Metallicity is routinely  estimated using the ratio [Fe/H]. The only previous estimate of Trumpler~20 metallicity from high-resolution spectroscopy is the
Platais et al. (2008) one. The spectrum of a single giant  (MG 675) is used to suggest for the whole cluster
metal abundance. This star  is not a clump star (see Platais et al. 2008, their Fig.~2), but most probably a bright red giant.
The resulting value from these authors is [Fe/H] = $-0.11\pm$0.13, indicating that Trumpler~20 is slightly metal-poor with respect to the Sun, 
although the value is compatible with solar, within the uncertainty.

Our spectroscopic campaign focusses on the cluster red giant clump, one magnitude fainter, which we sample  with eight stars (see Table~1).
Six out of eight turn out to be {\it bona fide} radial velocity members. For all these stars, except for star 496, which is a fast rotator, we
provide abundance analysis.
We  find an average [Fe/H] = 0.09$\pm$0.01, with no dispersion ($\sigma$=  0.02$\pm$0.01).  In this case, however, 0.01 is simply the internal dispersion
from the mean. Based on the analysis presented above (see Table~5), we derive a more reliable uncertainty of 0.074 when the errors on the
atmospheric parameters are  assumed as independent, or, in the opposite case, 0.095. We therefore adopt the value [Fe/H]=0.09$\pm$0.10.

While, compatible with the Platais et al. (2008) figure, our result 
suggests that Trumpler~20 is slightly more metal-rich than the Sun. We will discuss later the implication of this result on the cluster parameters.
This value of the metal abundance is not unexpected. Located well within the solar ring, Trumpler~20 is placed comfortably 
in the trend of abundance as a function of galactocentric distance. According to Magrini et al. (2009), the mean metallicity of 
old open clusters younger than $\sim$ 4 Gyr  is $\sim$ 0.1 dex at the galactocentric distance of Trumpler~20 ($\sim$ 7 kpc).

\subsection{$\alpha$ elements}
Trumpler~20 is an intermediate-age open cluster located inside the solar circle. It is therefore interesting to compare its chemical properties with stars
and star clusters located in the same region of the Milky Way.
The comparison of $\alpha$ element abundance ratios (for Mg, Ca, Ti, and Si)  with old open clusters and giant stars located in the inner disc  in shown in
Fig.~2. Giant stars  from Bensby et al.  (2010) are shown as grey circles, while the inner disc old open clusters NGC 6192, NGC 6404, and NGC 6583
from Magrini et al. (2010) are indicated as filled triangles. Our clump stars are shown as filled circles. 
The simultaneous inspection of this figure and Table~5 allows us to suggest that, overall, Trumpler~20 follows the trend of other inner disc indicators.
In detail, [Ca/Fe], [Ti/Fe], . and [Si/Fe] follow giant stars of similar metallicity very closely, namely their abundance ratios are nearly solar. 
The only marginally deviating element is Mg, which for the same [Fe/H], appears somewhat under-abundant, but still within the scatter. This also applies to the comparison with open clusters.

\begin{figure}
\includegraphics[width=\columnwidth]{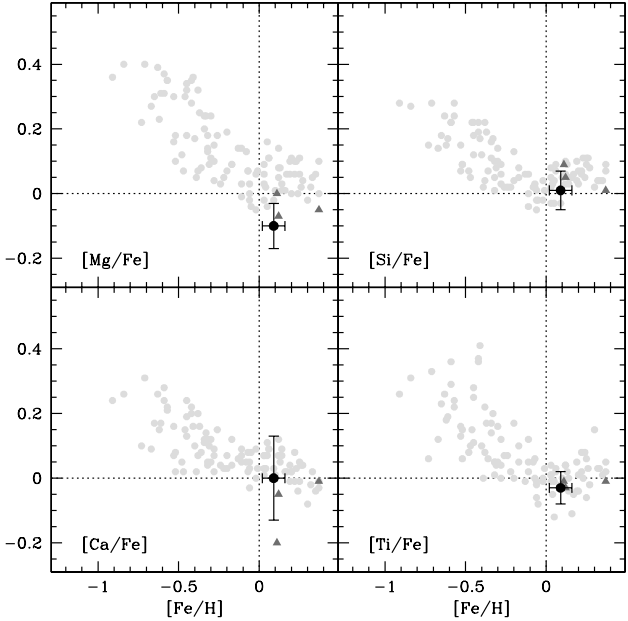}
\caption{Comparison between our $\alpha-$element ratios (black solid dots)  including error bars derived from Table~5, with data 
from old open clusters in the inner disc from Magrini et al. (2010; grey triangles) and inner disc giants from Bensby et al. (2010; light grey dots) }
\end{figure}

\subsection{Iron peak elements}
We compare two iron-peak element abundance ratios (Ni and  Cr) with the literature values in Fig.~3.
We could not compare V with thin and thick disc stars becuase the measure is based on a single line only, but this element ratio is clearly over-abundant with respect to
stars of the same metallicity. Hyperfine structure (HFS) effects can be the culprit, as amply discussed in Pancino et al. (2010). 
Cr (lower panel) and Ni (upper panel), nicely compare with disc stars, although the Reddy et al. (2003, 2006) samples do not contain as 
many stars with similar metallicity as Trumpler~20. Besides, it is reassuring to note that for these two elements we find the same trend shown by the inner disc
open cluster sample from Magrini et al. (2010, grey triangles).

\subsection{Neutron capture elements}
A recent review of neutron capture elements in intermediate age and old open clusters was presented by Mishenina et al. (2013), particularly for
barium (Ba) and yttrium (Y). This study demonstrates that Ba is routinely over-abundant with respect to the Sun, while Y is either solar or under-abundant with respect to the Sun.
In Fig.~4 we compare our findings  (filled black circles) for Ba and Y with thin (Reddy et al. 2003, large light-grey filled circles) and thick disc stars (Reddy et al. 2006, small light-grey filled circles), 
and with open cluster data from Mishenina et al. (2013, grey triangles). From this figure one can readily realize that Trumpler~20 follows the general trend. 
The value of [Ba/Fe] is marginally over-abundant ($\approx$ 0.20),
while Y is under-abundant. The same trend is visible in the sample of Pancino et al. (2010).

\begin{figure}
\includegraphics[width=\columnwidth]{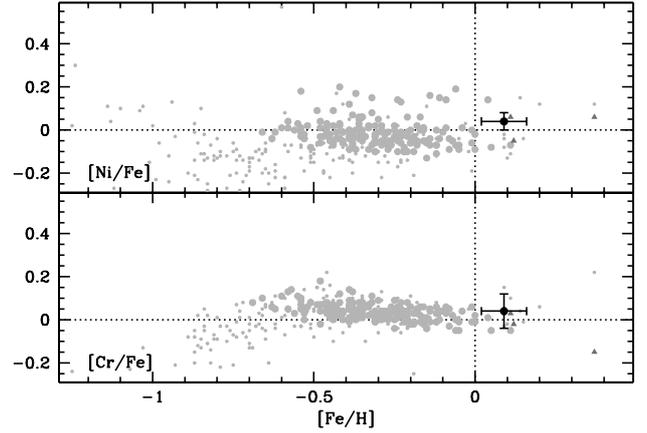}
\caption{ Mean V (lower panel), Cr (middle panel),  and Ni (upper panel) abundance ratios for Trumpler~20 giants (filled black circles), compared with thin disc stars (large light grey filled circles),
thick disc stars (small light grey symbols), and inner disc open clusters (Magrini et al. 2010, grey triangles)}
\end{figure}

\begin{figure}
\includegraphics[width=\columnwidth]{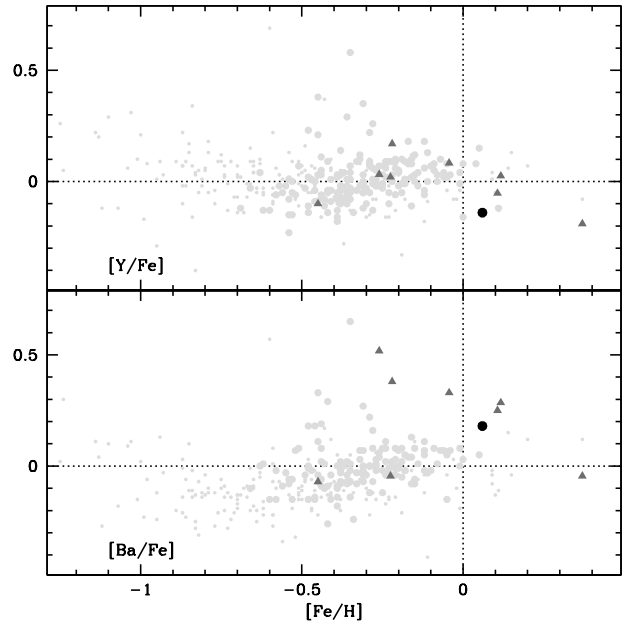}
\caption{Mean Ba (lower panel) and Y (upper panel) abundance ratios for Trumpler~20 giants (filled black circles), compared with thin disc stars (large light grey filled circles),
thick disc stars (small light grey symbols), and intermediate-age and old open clusters (grey triangles) }
\end{figure}

\begin{figure}
\includegraphics[width=\columnwidth]{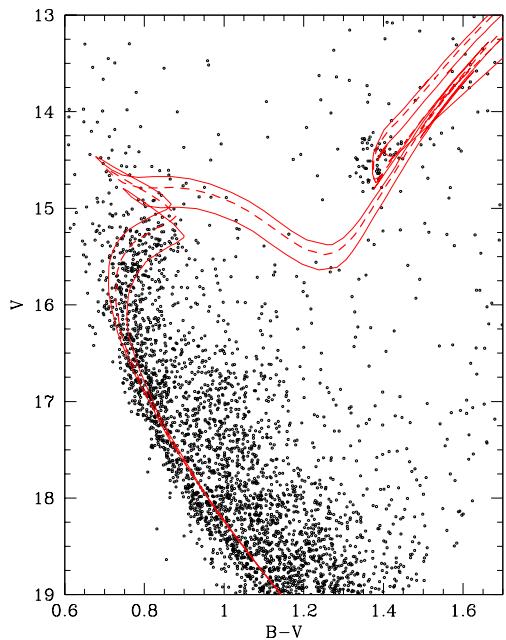}
\caption{Colour-magnitude diagram of Trumpler~20 from the photometric study of Carraro et al. (2010). Only stars within 6 arcmin from the cluster centre are considered. Over-imposed are three isochrones displaced using as  reddening
0.34 mag and  as apparent distance modulus 13.6 mag. The solid dashed isochrone is the best fit, for an age of 1.5 Gyr. To illustrate uncertainties in the age, we display two additional isochrones
(solid red symbols) for an age of 1.4 and 1.7 Gyr.  See text for details.
}
\end{figure}

\subsection{Revision of Trumpler~20 fundamental parameters}
Carraro et al. (2010) derived the fundamental parameters of Trumpler~20, and adopted solar metallicity. With the spectroscopic estimate of the metallicity obtained in this work,
we can now determine  the  fundamental parameters of the cluster more accurately.

The empirical metallicity [Fe/H] = +0.09 turns into the theoretical metallicity  Z=0.023 (see e.g. Carraro et al. 1999). We generated isochrones\footnote{\url{http://stev.oapd.inaf.it/cgi-bin/cmd}}
from the Padova suite of models \citep[][]{bressan12} for this metallicity and different ages. 
The best fit was obtained for an age of 1.5 Gyr, as illustrated in Fig.~5, where we also display two additional isochrones (for 1.4 and 1.7 Gyr) to illustrate fitting uncertainties.
The implied reddening is  $0.34\pm0.05$ mag and the apparent distance modulus $m-M  =13.6\pm0.2$  mag, where uncertainties are inferred by moving the isochrones
around the CMD. The fit of the turnoff region is reasonable in spite of the heavy field star contamination and the presence of a significant fraction of binary stars (Carraro et al. 2010).
As mentioned already in this last study, there is a marginal difficulty in matching  the mean colour of the RGB clump. This is quite common in the  literature (see, e.g., Carraro \& Costa 2007)
and can be ascribed to a variety of effects, the most important is  probably related to the isochrone transformation from the theoretical to the empirical plane and, also,
the calibration of the mixing length parameter. 

Because wack any indication of abnormal absorption along the line of sight to Trumpler 20, we adopt $R_{V}$= 3.1, and derive a heliocentric distance of 3.2 kpc, slightly larger
than the estimate  of Carraro et al. (2010). If we adopt 8.5 kpc as the Sun distance to the Galactic centre,
the galactic Cartesian coordinates for Trumpler 20 are:  $X  = 6.6$ kpc, $Y= -3.0$ kpc , and $Z = 120$ pc. 
As a consequence, the cluster distance to the Galactic Centre  $R_{GC}$  is 7.3$\pm$0.3 kpc.

\subsection{Trumpler~20 put into context}
Magrini et al. (2010) studied the radial abundance gradient in the inner galactic disc, and concluded that the inner disc gradient  (i.e. inside the solar circle)  is steeper than the
outer disc gradient.  This evidence has been recently confirmed by the new analysis of APOGEE data by Frinchaboy et al. (2013).
The addition of Trumpler~20 (see Fig.~6) confirms this piece of evidence. Located at 7.3 kpc from the Galactic centre and with a mean iron abundance [Fe/H] = +0.09, Trumpler~20
fits the trend defined by the other inner disc clusters. With the inclusion of Trumpler~20 the slope of the gradient is d[Fe/H]/dR = $-0.18\pm0.08$ dex/kpc.
One interesting aspect to mention is that most inner disc clusters seem to preferentially locate in the fourth Galactic quadrant (270 $\leq l \leq$ 360). In the sample  of Magrini et al. (2010)
only two clusters (NGC 6583 and NGC 6705) are located in the first Galactic quadrant. It would be interesting to extend the sample to more first quadrant clusters
to search for abundance inhomogeneity or azimuthal gradients (Villanova et al. 2005; L\'epine et al. 2011).

\begin{figure}
\includegraphics[width=\columnwidth]{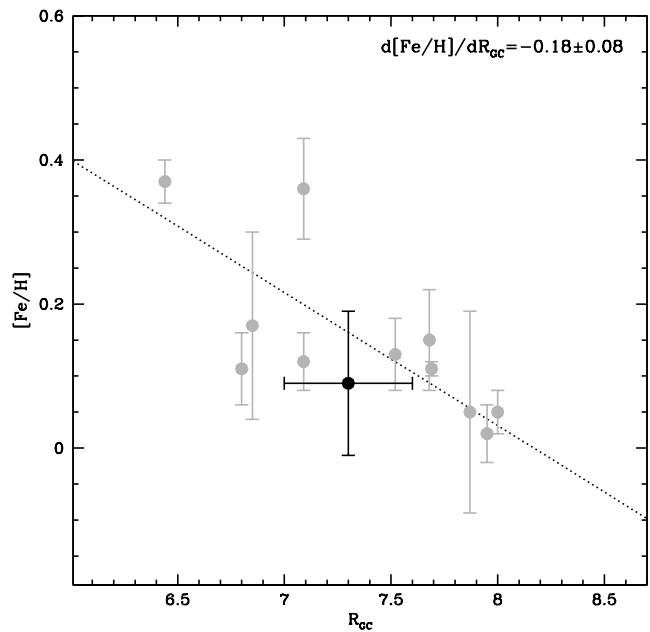}
\caption{The inner disc  radial abundance gradient from Magrini et al, (2010; grey symbols) with the addition of Trumpler~20 (black circle). The dotted line is a weighted
least square fit to the points, that yields $-0.18$ as value of the slope.}
\end{figure}

\section{Conclusions}
In this paper we have presented the first high-resolution spectroscopic study of a sample of red clump stars in the rich old open cluster, 
Trumpler~20.  The new  results fully support the findings in the previous photometric study by Carraro et al. (2010).\\

\noindent
Our results can be summarized as follows: 
\begin{itemize}
\item   the cluster metallicity is mildly super-solar, with [Fe/H]=+0.09$\pm$0.10. We did not detect any significant spread among the five {\it bona fide}
radial velocity members we were able to analyze;
\item   the abundance analysis reveals that Trumpler~20 is similar to other open clusters of the same age and metallicity as far as iron-peak  and neutron-capture
elements are concerned. The only deviation we found is for the $\alpha$-element Mg, which is under-abundant with respect to the Sun.  Its value is, however, consistent
with the region covered by inner disc clusters or in inner disc giant stars of similar metallicity;
\item   we have revised the cluster fundamental parameters. The new values for reddening and distance modulus  are  0.34$\pm$0.05 mag  and 13.6$\pm$0.2 mag, respectively.
By using these figures, we derive a cluster galacto-centric distance of 7.3 kpc;
\item  at this distance, and with a metallicity of [Fe/H] =+0.09, Trumpler~20 nicely matches the radial abundance gradient of the inner Galactic disc.
\item   we did not detect any spread in the Na abundances in our limited sample. More data would be required, however, to firmly discard the 
presence of multiple stellar populations in the cluster (Platais et al. 2012).
\end{itemize}


Further studies of this cluster should concentrate on estimating the binary fraction, and refining the shape of the main sequence via a detailed membership analysis.
Quantifying the fraction of interlopers is an important step toward the determination of the cluster mass.

\begin{acknowledgements}
J. Ahumada is grateful to ESO for supporting his visit to the Santiago premises in 2011 and 2012, where this project started. 
S. Villanova gratefully acknowledges the support provided by FONDECYT N. 1130721.
\end{acknowledgements}

\begin{table*} 
\caption{Adopted line list, atomic parameters, equivalent widths (EW), and abundances (X).} 
\begin{tabular}{rrrrrrrrrrrrrr}
\hline\hline
 $\lambda$ & Element & $\chi$ & log(gf) & EW & X & EW & X & EW& X & EW & X& EW & X\\
\hline
\hline
&&&& 207 && 258 && 542 && 609 && 80 &\\
6154.219& 11.0& 2.100& -1.600& 82.3& 6.60& 77.7& 6.52& 81.8& 6.61& 84.8& 6.68& 86.5& 6.71\\ 
6160.742& 11.0& 2.104& -1.260& 107.1& 6.67& 104.8& 6.63& 105.4& 6.67& 97.5& 6.56& 93.8& 6.50\\
5711.083& 12.0& 4.340& -1.670& 125.4& 7.49& 133.7& 7.62& 125.8& 7.52& 128.3& 7.57& 125.4& 7.53\\ 
6318.702& 12.0& 5.110& -1.950& 76.0& 7.76& 0.0& 0.00& 0.0& 0.00& 0.0& 0.00& 0.0& 0.00\\
6696.014& 13.0& 3.140& -1.562& 0.0& 0.00& 76.6& 6.72& 73.4& 6.68& 75.6& 6.74& 69.0& 6.63\\
6698.663& 13.0& 3.140& -1.830& 49.2& 6.53& 43.2& 6.42& 52.3& 6.60& 44.4& 6.48& 41.0& 6.42\\
5517.529& 14.0& 5.082& -2.554& 30.4& 7.77& 0.0& 0.00& 29.6& 7.77& 0.0& 0.00& 0.0& 0.00\\
5645.603& 14.0& 4.930& -2.120& 64.7& 7.83& 0.0& 0.00& 0.0& 0.00& 61.6& 7.80& 64.0& 7.82\\
5684.479& 14.0& 4.930& -1.700& 78.6& 7.65& 84.4& 7.74& 79.5& 7.68& 73.0& 7.57& 83.7& 7.74\\
5690.419& 14.0& 4.930& -1.840& 67.4& 7.60& 62.5& 7.49& 69.5& 7.65& 67.0& 7.61& 73.8& 7.72\\
5701.098& 14.0& 4.930& -2.080& 58.0& 7.66& 60.5& 7.69& 60.0& 7.71& 60.2& 7.73& 56.2& 7.63\\
6125.014& 14.0& 5.610& -1.580& 49.4& 7.72& 51.0& 7.74& 50.0& 7.74& 52.7& 7.80& 45.8& 7.66\\
6142.481& 14.0& 5.619& -1.530& 47.7& 7.67& 48.1& 7.66& 48.9& 7.70& 42.0& 7.57& 46.3& 7.64\\
6145.010& 14.0& 5.610& -1.450& 48.6& 7.59& 52.3& 7.65& 50.6& 7.64& 55.8& 7.74& 49.6& 7.61\\
6244.465& 14.0& 5.616& -1.340& 63.0& 7.76& 69.3& 7.86& 71.3& 7.91& 69.2& 7.88& 69.1& 7.86\\
5260.377& 20.0& 2.520& -1.820& 59.2& 6.33& 0.0& 0.00& 65.8& 6.50& 66.3& 6.53& 58.6& 6.37\\
5349.458& 20.0& 2.709& -0.178& 0.0& 0.00& 123.9& 6.38& 132.1& 6.57& 123.8& 6.43& 124.5& 6.45\\
5867.554& 20.0& 2.930& -1.630& 51.4& 6.42& 54.4& 6.47& 55.8& 6.52& 55.5& 6.54& 53.9& 6.51\\
6161.287& 20.0& 2.523& -1.293& 107.8& 6.68& 111.3& 6.74& 114.1& 6.82& 111.5& 6.80& 0.0& 0.00\\
6166.429& 20.0& 2.521& -1.136& 108.1& 6.52& 110.1& 6.56& 108.3& 6.55& 96.8& 6.36& 97.0& 6.37\\
6455.593& 20.0& 2.520& -1.320& 100.0& 6.55& 102.6& 6.60& 100.4& 6.59& 99.2& 6.60& 96.3& 6.54\\
6572.774& 20.0& 0.000& -4.390& 122.8& 6.26& 113.2& 6.06& 116.5& 6.21& 110.5& 6.18& 113.1& 6.22\\
5062.094& 22.0& 2.160& -0.420& 45.2& 4.82& 0.0& 0.00& 56.7& 5.07& 0.0& 0.00& 43.5& 4.85\\
5113.437& 22.0& 1.440& -0.820& 0.0& 0.00& 71.5& 4.92& 0.0& 0.00& 0.0& 0.00& 0.0& 0.00\\
5145.457& 22.0& 1.460& -0.600& 83.0& 5.00& 85.5& 5.04& 76.3& 4.88& 84.6& 5.12& 80.7& 5.02\\
5282.413& 22.0& 1.053& -1.640& 60.6& 5.01& 65.0& 5.09& 66.3& 5.17& 68.4& 5.25& 69.0& 5.27\\
5295.774& 22.0& 1.050& -1.680& 53.3& 4.91& 0.0& 0.00& 0.0& 0.00& 52.1& 4.96& 55.1& 5.02\\
5490.144& 22.0& 1.460& -0.950& 65.1& 4.88& 68.0& 4.93& 64.5& 4.91& 67.8& 5.01& 0.0& 0.00\\
5648.560& 22.0& 2.495& -0.350& 35.6& 4.91& 0.0& 0.00& 0.0& 0.00& 36.8& 4.99& 0.0& 0.00\\
5739.465& 22.0& 2.249& -0.724& 30.3& 4.89& 37.4& 5.02& 33.7& 4.99& 0.0& 0.00& 0.0& 0.00\\
5922.105& 22.0& 1.050& -1.450& 78.6& 5.10& 73.7& 4.99& 75.8& 5.09& 75.0& 5.11& 73.0& 5.06\\
5965.822& 22.0& 1.870& -0.540& 71.7& 5.05& 69.5& 4.99& 71.9& 5.09& 79.9& 5.29& 72.5& 5.14\\
6091.167& 22.0& 2.270& -0.430& 55.2& 5.07& 53.5& 5.03& 49.5& 5.00& 0.0& 0.00& 0.0& 0.00\\
6126.214& 22.0& 1.070& -1.360& 82.0& 5.07& 82.7& 5.07& 78.9& 5.06& 76.3& 5.04& 71.6& 4.94\\
6258.098& 22.0& 1.440& -0.340& 103.1& 4.93& 99.1& 4.84& 104.7& 5.01& 98.7& 4.93& 95.2& 4.85\\
6261.094& 22.0& 1.430& -0.440& 124.6& 5.47& 121.5& 5.39& 123.3& 5.48& 118.0& 5.42& 114.1& 5.34\\
6554.220& 22.0& 1.440& -1.210& 77.1& 5.22& 0.0& 0.00& 77.2& 5.26& 76.6& 5.30& 70.7& 5.18\\
5211.523& 22.1& 2.590& -1.456& 63.1& 5.07& 60.8& 4.99& 66.8& 5.19& 0.0& 0.00& 0.0& 0.00\\
5336.783& 22.1& 1.582& -1.600& 104.2& 4.90& 108.5& 4.96& 105.9& 4.96& 98.0& 4.83& 96.5& 4.77\\
5418.762& 22.1& 1.580& -2.080& 75.6& 4.78& 83.6& 4.92& 75.3& 4.80& 73.2& 4.78& 73.0& 4.74\\
5490.688& 22.1& 1.566& -2.730& 0.0& 0.00& 0.0& 0.00& 0.0& 0.00& 61.6& 5.19& 0.0& 0.00\\
5670.847& 23.0& 1.081& -0.578& 94.0& 4.65& 89.1& 4.53& 84.4& 4.47& 82.8& 4.48& 84.6& 4.52\\
4801.021& 24.0& 3.120& -0.140& 73.3& 5.53& 76.0& 5.60& 85.6& 5.86& 0.0& 0.00& 0.0& 0.00\\
4936.335& 24.0& 3.113& -0.280& 77.2& 5.77& 61.6& 5.38& 72.3& 5.67& 79.4& 5.90& 72.4& 5.71\\
4964.923& 24.0& 0.940& -2.490& 89.4& 5.68& 0.0& 0.00& 0.0& 0.00& 0.0& 0.00& 0.0& 0.00\\
5214.125& 24.0& 3.369& -0.753& 0.0& 0.00& 42.1& 5.69& 41.6& 5.71& 0.0& 0.00& 0.0& 0.00\\
5238.954& 24.0& 2.710& -1.410& 48.5& 5.73& 54.1& 5.83& 0.0& 0.00& 43.2& 5.66& 0.0& 0.00\\
 \hline 
\end{tabular} 
\end{table*} 

\addtocounter{table}{-1} 
\begin{table*} 
\caption{Adopted line list, atomic parameters, equivalent widths (EW), and abundances (X) .}
\begin{tabular}{rrrrrrrrrrrrrr}
\hline\hline
$\lambda$& Element& $\chi$& log (gf) & EW& X& EW& X& EW& X& EW& X& EW & X\\
\hline
&&&& 207 && 258 && 542 && 609 && 80 &\\
\hline
5247.564& 24.0& 0.961& -1.550& 121.0& 5.44& 138.5& 5.79& 126.9& 5.61& 0.0& 0.00& 124.8& 5.61\\
5271.993& 24.0& 3.449& -0.482& 58.5& 5.86& 0.0& 0.00& 0.0& 0.00& 0.0& 0.00& 0.0& 0.00\\
5287.170& 24.0& 3.438& -0.957& 34.1& 5.80& 0.0& 0.00& 40.9& 5.97& 39.2& 5.95& 31.6& 5.78\\
5296.691& 24.0& 0.983& -1.310& 132.2& 5.44& 142.8& 5.63& 140.8& 5.65& 137.3& 5.62& 136.2& 5.60\\
5300.743& 24.0& 0.980& -2.120& 112.6& 5.82& 106.8& 5.68& 113.7& 5.89& 108.2& 5.81& 106.0& 5.76\\
5304.173& 24.0& 3.460& -0.730& 0.0& 0.00& 38.6& 5.69& 0.0& 0.00& 44.5& 5.85& 0.0& 0.00\\
5318.764& 24.0& 3.440& -0.720& 42.1& 5.73& 0.0& 0.00& 49.0& 5.90& 0.0& 0.00& 0.0& 0.00\\
5329.137& 24.0& 2.914& -0.194& 120.0& 6.30& 115.2& 6.22& 0.0& 0.00& 0.0& 0.00& 125.8& 6.47\\
5348.315& 24.0& 1.004& -1.140& 142.1& 5.46& 147.6& 5.55& 147.5& 5.59& 148.5& 5.65& 146.1& 5.61\\
5628.635& 24.0& 3.420& -0.790& 0.0& 0.00& 0.0& 0.00& 43.8& 5.82& 0.0& 0.00& 0.0& 0.00\\
6330.086& 24.0& 0.940& -2.880& 92.3& 5.82& 91.1& 5.79& 88.3& 5.79& 85.4& 5.77& 83.3& 5.73\\
5058.490& 26.0& 3.642& -2.750& 0.0& 0.00& 0.0& 0.00& 0.0& 0.00& 39.9& 7.61& 40.3& 7.61\\
5141.737& 26.0& 2.424& -2.124& 0.0& 0.00& 126.5& 7.55& 117.9& 7.42& 0.0& 0.00& 129.6& 7.69\\
5180.060& 26.0& 4.473& -1.120& 77.9& 7.74& 0.0& 0.00& 80.0& 7.81& 0.0& 0.00& 76.1& 7.73\\
5196.056& 26.0& 4.256& -0.590& 99.9& 7.52& 109.2& 7.73& 107.3& 7.71& 104.2& 7.68& 104.7& 7.68\\
5197.934& 26.0& 4.301& -1.480& 63.5& 7.56& 73.7& 7.80& 71.9& 7.78& 67.3& 7.70& 77.1& 7.93\\
5217.385& 26.0& 3.211& -1.020& 0.0& 0.00& 0.0& 0.00& 0.0& 0.00& 139.5& 7.46& 0.0& 0.00\\
5223.177& 26.0& 3.635& -2.243& 59.2& 7.47& 64.2& 7.56& 69.0& 7.72& 62.3& 7.59& 66.0& 7.68\\
5225.525& 26.0& 0.110& -4.669& 137.7& 7.52& 151.6& 7.75& 154.7& 7.87& 0.0& 0.00& 133.8& 7.54\\
5242.488& 26.0& 3.634& -0.807& 0.0& 0.00& 115.2& 7.40& 114.4& 7.41& 0.0& 0.00& 117.5& 7.50\\
5243.771& 26.0& 4.256& -0.930& 79.1& 7.33& 90.6& 7.61& 0.0& 0.00& 93.8& 7.74& 78.7& 7.37\\
5247.048& 26.0& 0.087& -4.800& 0.0& 0.00& 128.4& 7.39& 135.0& 7.61& 134.7& 7.66& 0.0& 0.00\\
5253.015& 26.0& 2.279& -3.849& 0.0& 0.00& 70.9& 7.72& 62.9& 7.59& 61.2& 7.59& 56.3& 7.46\\
5253.459& 26.0& 3.283& -1.523& 0.0& 0.00& 104.0& 7.36& 105.2& 7.43& 102.2& 7.39& 111.2& 7.57\\
5288.524& 26.0& 3.694& -1.550& 98.0& 7.81& 92.9& 7.67& 95.1& 7.77& 84.3& 7.53& 88.3& 7.62\\
5294.540& 26.0& 3.640& -2.680& 43.7& 7.56& 41.7& 7.50& 46.9& 7.65& 34.9& 7.41& 0.0& 0.00\\
5295.310& 26.0& 4.415& -1.530& 0.0& 0.00& 54.4& 7.51& 47.4& 7.39& 53.3& 7.54& 51.2& 7.48\\
5321.105& 26.0& 4.434& -1.261& 72.5& 7.69& 75.4& 7.76& 66.1& 7.57& 60.8& 7.47& 61.2& 7.47\\
5326.140& 26.0& 3.573& -2.210& 79.2& 7.84& 70.7& 7.61& 67.1& 7.56& 0.0& 0.00& 71.4& 7.69\\
5386.327& 26.0& 4.154& -1.700& 63.7& 7.60& 54.0& 7.37& 58.2& 7.50& 0.0& 0.00& 51.0& 7.35\\
5401.260& 26.0& 4.320& -1.720& 63.7& 7.81& 56.0& 7.62& 64.6& 7.85& 54.9& 7.65& 0.0& 0.00\\
5441.333& 26.0& 4.312& -1.590& 63.2& 7.66& 50.7& 7.37& 56.9& 7.54& 56.8& 7.56& 56.6& 7.54\\
5460.870& 26.0& 3.071& -3.530& 29.0& 7.43& 26.7& 7.36& 31.3& 7.50& 35.4& 7.61& 28.8& 7.46\\
5461.543& 26.0& 4.445& -1.612& 55.7& 7.65& 49.6& 7.51& 58.0& 7.73& 58.1& 7.75& 47.0& 7.49\\
5464.275& 26.0& 4.143& -1.582& 76.6& 7.76& 79.5& 7.82& 80.8& 7.88& 0.0& 0.00& 72.2& 7.70\\
5470.086& 26.0& 4.446& -1.610& 55.2& 7.64& 53.6& 7.59& 52.0& 7.59& 46.6& 7.49& 50.0& 7.56\\
5494.458& 26.0& 4.070& -1.960& 60.7& 7.70& 64.4& 7.77& 0.0& 0.00& 64.4& 7.84& 55.5& 7.62\\
5522.444& 26.0& 4.210& -1.400& 78.5& 7.70& 66.5& 7.40& 68.8& 7.50& 76.5& 7.70& 73.2& 7.61\\
5525.540& 26.0& 4.230& -1.184& 76.7& 7.47& 85.1& 7.66& 78.9& 7.55& 83.5& 7.69& 78.9& 7.57\\
5538.512& 26.0& 4.218& -1.559& 64.7& 7.54& 72.7& 7.72& 65.6& 7.59& 68.5& 7.69& 77.0& 7.88\\
5539.276& 26.0& 3.640& -2.610& 42.0& 7.43& 47.5& 7.54& 50.1& 7.63& 44.3& 7.53& 50.2& 7.65\\
5546.503& 26.0& 4.371& -1.080& 0.0& 0.00& 83.9& 7.68& 80.0& 7.62& 82.9& 7.72& 86.8& 7.80\\
5552.685& 26.0& 4.950& -1.800& 0.0& 0.00& 0.0& 0.00& 19.1& 7.57& 0.0& 0.00& 0.0& 0.00\\
5560.208& 26.0& 4.430& -1.000& 72.4& 7.40& 65.4& 7.23& 76.9& 7.53& 70.0& 7.40& 73.2& 7.47\\
5576.087& 26.0& 3.430& -0.840& 0.0& 0.00& 149.2& 7.52& 140.9& 7.44& 136.4& 7.40& 135.2& 7.39\\
5577.018& 26.0& 5.030& -1.520& 0.0& 0.00& 0.0& 0.00& 0.0& 0.00& 0.0& 0.00& 43.6& 7.98\\
5587.568& 26.0& 4.140& -1.650& 73.8& 7.76& 0.0& 0.00& 0.0& 0.00& 0.0& 0.00& 0.0& 0.00\\
5608.969& 26.0& 4.210& -2.360& 0.0& 0.00& 33.1& 7.63& 36.4& 7.73& 25.2& 7.49& 30.5& 7.61\\
5611.351& 26.0& 3.630& -2.960& 37.8& 7.68& 36.6& 7.64& 33.0& 7.60& 0.0& 0.00& 31.6& 7.59\\
5618.627& 26.0& 4.210& -1.260& 84.7& 7.70& 68.6& 7.31& 81.5& 7.65& 79.4& 7.63& 72.4& 7.45\\
5619.594& 26.0& 4.390& -1.480& 71.5& 7.81& 65.1& 7.65& 67.0& 7.73& 65.9& 7.73& 62.5& 7.64\\
5635.816& 26.0& 4.260& -1.560& 55.3& 7.37& 55.9& 7.37& 57.9& 7.45& 62.7& 7.58& 56.2& 7.42\\
5636.695& 26.0& 3.640& -2.530& 45.0& 7.41& 53.4& 7.57& 47.7& 7.49& 45.1& 7.46& 0.0& 0.00\\
5638.255& 26.0& 4.220& -0.690& 105.3& 7.59& 97.3& 7.41& 109.8& 7.71& 112.0& 7.77& 102.8& 7.58\\
5649.981& 26.0& 5.100& -0.750& 0.0& 0.00& 59.1& 7.59& 0.0& 0.00& 0.0& 0.00& 0.0& 0.00\\
5651.462& 26.0& 4.470& -1.780& 35.9& 7.41& 42.5& 7.55& 40.0& 7.52& 37.9& 7.49& 38.4& 7.50\\
5652.312& 26.0& 4.260& -1.760& 53.5& 7.53& 59.0& 7.64& 55.4& 7.60& 54.4& 7.60& 50.6& 7.51\\
5677.681& 26.0& 4.100& -2.680& 0.0& 0.00& 0.0& 0.00& 0.0& 0.00& 0.0& 0.00& 27.9& 7.74\\
5679.019& 26.0& 4.652& -0.680& 81.4& 7.51& 79.0& 7.46& 81.4& 7.54& 78.5& 7.50& 77.3& 7.46\\
5680.234& 26.0& 4.190& -2.370& 31.4& 7.58& 38.5& 7.73& 35.6& 7.70& 32.3& 7.64& 0.0& 0.00\\
5701.542& 26.0& 2.560& -2.130& 135.7& 7.75& 136.8& 7.75& 130.7& 7.69& 131.2& 7.74& 134.8& 7.80\\
\hline 
\end{tabular} 
\end{table*} 

\addtocounter{table}{-1} 
\begin{table*} 
\caption{Adopted line list, atomic parameters, equivalent widths (EW), and abundances (X).}
\begin{tabular}{ccrrrrrrrrrrrr}
\hline\hline
$\lambda$& Element & $\chi$i& log(gf) & EW& X& EW& X& EW& X& EW& X& EW& X\\
\hline
&&&& 207 && 258 && 542 && 609 && 80 &\\
\hline
5705.460& 26.0& 4.301& -1.455& 67.1& 7.58& 72.4& 7.68& 69.5& 7.65& 74.4& 7.80& 70.6& 7.70\\
5717.827& 26.0& 4.280& -0.990& 96.8& 7.76& 96.8& 7.76& 98.0& 7.81& 92.1& 7.71& 97.4& 7.82\\
5731.758& 26.0& 4.260& -1.060& 84.3& 7.53& 90.7& 7.67& 88.1& 7.65& 0.0& 0.00& 82.0& 7.52\\
5741.844& 26.0& 4.260& -1.640& 59.8& 7.54& 0.0& 0.00& 61.3& 7.60& 58.1& 7.56& 63.1& 7.66\\
5742.953& 26.0& 4.180& -2.320& 38.0& 7.66& 0.0& 0.00& 46.3& 7.86& 43.4& 7.82& 0.0& 0.00\\
5752.028& 26.0& 4.549& -0.867& 83.3& 7.61& 73.2& 7.39& 86.0& 7.70& 76.3& 7.51& 72.4& 7.41\\
5753.120& 26.0& 4.260& -0.588& 100.0& 7.40& 107.9& 7.57& 107.4& 7.58& 99.2& 7.44& 103.6& 7.52\\
5760.342& 26.0& 3.640& -2.450& 60.0& 7.64& 0.0& 0.00& 54.7& 7.55& 63.2& 7.77& 47.4& 7.41\\
5775.076& 26.0& 4.220& -1.040& 84.4& 7.46& 91.5& 7.60& 0.0& 0.00& 87.5& 7.58& 0.0& 0.00\\
5853.147& 26.0& 1.480& -5.170& 0.0& 0.00& 0.0& 0.00& 48.0& 7.59& 53.6& 7.73& 51.3& 7.68\\
5855.076& 26.0& 4.610& -1.540& 43.7& 7.49& 44.1& 7.49& 51.2& 7.67& 50.2& 7.67& 46.8& 7.58\\
5856.086& 26.0& 4.290& -1.550& 62.5& 7.55& 72.8& 7.78& 65.8& 7.65& 72.5& 7.84& 66.0& 7.67\\
5859.583& 26.0& 4.549& -0.388& 94.7& 7.41& 98.1& 7.48& 99.8& 7.55& 96.1& 7.50& 93.0& 7.42\\
5861.101& 26.0& 4.280& -2.400& 0.0& 0.00& 25.7& 7.56& 30.2& 7.70& 22.3& 7.52& 29.8& 7.70\\
5862.353& 26.0& 4.549& -0.148& 102.6& 7.34& 103.0& 7.35& 106.0& 7.43& 107.2& 7.48& 99.7& 7.32\\
5881.273& 26.0& 4.610& -1.770& 41.4& 7.67& 38.1& 7.59& 0.0& 0.00& 0.0& 0.00& 0.0& 0.00\\
5905.668& 26.0& 4.650& -0.700& 85.7& 7.62& 84.8& 7.59& 79.3& 7.50& 75.1& 7.43& 82.2& 7.58\\
5916.246& 26.0& 2.453& -2.857& 0.0& 0.00& 112.2& 7.80& 106.8& 7.74& 0.0& 0.00& 89.1& 7.35\\
5927.785& 26.0& 4.650& -1.020& 66.6& 7.51& 63.1& 7.42& 66.0& 7.51& 60.9& 7.42& 66.6& 7.54\\
5929.672& 26.0& 4.550& -1.170& 69.4& 7.60& 73.2& 7.68& 72.2& 7.69& 66.8& 7.60& 70.5& 7.67\\
5930.179& 26.0& 4.650& -0.010& 112.7& 7.46& 120.1& 7.59& 117.8& 7.57& 108.9& 7.43& 112.8& 7.50\\
5934.652& 26.0& 3.928& -1.060& 104.1& 7.51& 103.6& 7.50& 102.7& 7.51& 99.6& 7.48& 107.8& 7.63\\
5956.693& 26.0& 0.860& -4.500& 114.3& 7.53& 112.7& 7.47& 113.0& 7.55& 109.2& 7.52& 116.0& 7.66\\
5983.675& 26.0& 4.549& -0.558& 97.3& 7.62& 91.3& 7.48& 92.8& 7.54& 94.7& 7.61& 0.0& 0.00\\
6003.010& 26.0& 3.880& -0.970& 108.3& 7.45& 108.6& 7.46& 110.7& 7.53& 107.0& 7.48& 106.7& 7.47\\
6007.958& 26.0& 4.650& -0.620& 89.8& 7.64& 0.0& 0.00& 90.6& 7.68& 86.1& 7.61& 85.0& 7.57\\
6008.555& 26.0& 3.880& -0.830& 121.2& 7.55& 110.5& 7.36& 113.3& 7.44& 106.9& 7.35& 110.3& 7.41\\
6024.055& 26.0& 4.548& 0.200& 125.8& 7.37& 123.3& 7.33& 134.0& 7.51& 124.9& 7.40& 119.1& 7.30\\
6027.049& 26.0& 4.070& -1.020& 91.5& 7.41& 94.2& 7.46& 89.5& 7.39& 93.6& 7.52& 98.2& 7.61\\
6056.001 26.0& 4.730& -0.340& 90.6& 7.44& 93.0& 7.48& 92.0& 7.49& 89.4& 7.46& 87.5& 7.41\\
6065.482 26.0& 2.610& -1.400& 0.0& 0.00& 165.9& 7.42& 167.9& 7.48& 165.9& 7.48& 163.3& 7.45\\
6079.004 26.0& 4.650& -0.960& 71.3& 7.54& 71.9& 7.55& 71.4& 7.57& 62.1& 7.38& 69.3& 7.53\\
6082.706 26.0& 2.220& -3.550& 92.7& 7.74& 88.9& 7.64& 0.0& 0.00& 0.0& 0.00& 79.6& 7.51\\
6093.639 26.0& 4.610& -1.340& 55.2& 7.52& 61.6& 7.66& 54.6& 7.53& 50.3& 7.46& 56.3& 7.58\\
6094.367 26.0& 4.650& -1.610& 46.7& 7.66& 44.1& 7.59& 47.7& 7.70& 42.7& 7.61& 42.2& 7.59\\
6096.661 26.0& 3.980& -1.810& 73.2& 7.65& 67.1& 7.51& 72.6& 7.66& 70.9& 7.66& 67.8& 7.58\\
6098.240 26.0& 4.560& -1.800& 0.0& 0.00& 39.4& 7.58& 39.2& 7.61& 43.7& 7.72& 0.0& 0.00\\
6120.244 26.0& 0.910& -5.930& 40.3& 7.45& 43.3& 7.48& 0.0& 0.00& 0.0& 0.00& 39.4& 7.50\\
6127.902 26.0& 4.143& -1.349& 86.6& 7.71& 80.1& 7.55& 77.6& 7.53& 86.7& 7.78& 81.9& 7.65\\
6151.613 26.0& 2.180& -3.230& 99.3& 7.51& 95.2& 7.40& 95.5& 7.47& 96.6& 7.54& 90.7& 7.39\\
6157.723 26.0& 4.070& -1.120& 99.6& 7.67& 0.0& 0.00& 107.7& 7.87& 96.8& 7.67& 95.9& 7.64\\
6159.369 26.0& 4.607& -1.890& 33.4& 7.60& 34.3& 7.61& 38.6& 7.73& 29.1& 7.53& 28.4& 7.51\\
6165.355 26.0& 4.140& -1.430& 78.9& 7.60& 79.5& 7.61& 78.6& 7.63& 76.7& 7.62& 73.1& 7.52\\
6173.331 26.0& 2.220& -2.800& 120.5& 7.61& 122.7& 7.64& 118.0& 7.60& 117.4& 7.63& 116.4& 7.60\\
6187.394 26.0& 2.830& -4.190& 32.4& 7.83& 0.0& 0.00& 0.0& 0.00& 0.0& 0.00& 0.0& 0.00\\
6187.985 26.0& 3.940& -1.660& 89.2& 7.79& 83.3& 7.65& 87.0& 7.77& 85.1& 7.76& 72.5& 7.47\\
6200.311 26.0& 2.610& -2.300& 118.7& 7.53& 117.6& 7.49& 120.3& 7.60& 116.8& 7.57& 117.0& 7.57\\
6213.427 26.0& 2.220& -2.450& 138.4& 7.60& 125.4& 7.34& 138.1& 7.64& 127.5& 7.48& 129.0& 7.50\\
6219.279 26.0& 2.200& -2.340& 148.5& 7.64& 143.5& 7.55& 144.6& 7.62& 138.4& 7.55& 143.7& 7.63\\
6220.776 26.0& 3.880& -2.370& 55.5& 7.70& 0.0& 0.00& 53.6& 7.69& 52.0& 7.68& 0.0& 0.00\\
6226.729 26.0& 3.880& -2.110& 59.2& 7.52& 59.0& 7.50& 59.0& 7.54& 52.3& 7.43& 56.1& 7.49\\
6232.638 26.0& 3.650& -1.180& 116.1& 7.51& 120.1& 7.58& 124.5& 7.69& 122.3& 7.68& 107.2& 7.40\\
6240.643 26.0& 2.220& -3.230& 103.8& 7.65& 96.3& 7.47& 98.0& 7.56& 93.9& 7.52& 101.1& 7.67\\
6246.315 26.0& 3.602& -0.778& 144.5& 7.49& 142.4& 7.46& 147.6& 7.56& 144.3& 7.53& 142.3& 7.51\\
\hline 
\end{tabular} 
\end{table*} 

\addtocounter{table}{-1} 
\begin{table*} 
\caption{Adopted line list, atomic parameters, equivalent widths (EW), and abundances (X)}
\begin{tabular}{ccrrrrrrrrrrrr}
\hline\hline
$\lambda$& Element& $\chi$& log(gf)& EW& X& EW& X& EW& X& EW& X& EW& X\\
\hline
&&&& 207 && 258 && 542 && 609 && 80 &\\
\hline
6252.556 &26.0& 2.400& -1.640& 171.7& 7.46& 162.2& 7.34& 0.0& 0.00& 0.0& 0.00& 0.0& 0.00\\
6265.131 &26.0& 2.180& -2.460& 146.1& 7.69& 146.1& 7.68& 152.9& 7.84& 148.4& 7.81& 143.8& 7.72\\
6270.221 &26.0& 2.860& -2.510& 98.5& 7.59& 106.7& 7.76& 100.6& 7.67& 100.4& 7.71& 102.3& 7.74\\
6271.272 &26.0& 3.332& -2.763& 66.6& 7.68& 59.3& 7.52& 64.6& 7.67& 0.0& 0.00& 66.0& 7.72\\
6297.789 &26.0& 2.220& -2.660& 132.2& 7.68& 123.4& 7.49& 126.8& 7.61& 124.0& 7.60& 122.9& 7.57\\
6302.492 &26.0& 3.686& -1.083& 118.4& 7.52& 110.1& 7.35& 0.0& 0.00& 115.5& 7.52& 110.7& 7.42\\
6311.494 &26.0& 2.830& -3.150& 0.0& 0.00& 0.0& 0.00& 0.0& 0.00& 75.8& 7.74& 0.0& 0.00\\
6315.814 &26.0& 4.070& -1.650& 74.3& 7.62& 70.4& 7.52& 79.6& 7.77& 70.8& 7.60& 74.2& 7.66\\
6322.684 &26.0& 2.590& -2.330& 112.4& 7.37& 120.4& 7.53& 118.4& 7.54& 111.2& 7.43& 105.0& 7.28\\
6330.842 &26.0& 4.730& -1.170& 63.1& 7.65& 0.0& 0.00& 55.2& 7.50& 0.0& 0.00& 54.6& 7.49\\
6335.327 &26.0& 2.200& -2.200& 151.7& 7.53& 156.2& 7.59& 150.0& 7.54& 145.4& 7.51& 152.6& 7.61\\
6336.821 &26.0& 3.686& -0.806& 144.4& 7.59& 138.9& 7.52& 138.4& 7.54& 139.2& 7.57& 141.6& 7.60\\
6380.738 &26.0& 4.190& -1.280& 86.0& 7.62& 89.1& 7.68& 85.8& 7.65& 90.4& 7.78& 0.0& 0.00\\
6392.533 &26.0& 2.280& -3.956& 64.4& 7.56& 65.4& 7.57& 56.4& 7.44& 55.0& 7.44& 0.0& 0.00\\
6416.918 &26.0& 4.795& -0.995& 57.0& 7.40& 55.2& 7.36& 53.2& 7.34& 55.4& 7.41& 51.4& 7.32\\
6430.844 &26.0& 2.176& -1.996& 177.6& 7.62& 0.0& 0.00& 0.0& 0.00& 0.0& 0.00& 173.5& 7.64\\
6436.399 &26.0& 4.190& -2.410& 38.2& 7.74& 32.6& 7.60& 0.0& 0.00& 34.9& 7.71& 37.4& 7.75\\
6464.664 &26.0& 0.958& -5.511& 61.3& 7.44& 56.9& 7.34& 56.0& 7.39& 55.3& 7.42& 54.7& 7.40\\
6481.868 &26.0& 2.280& -2.890& 121.5& 7.71& 124.9& 7.77& 111.1& 7.54& 117.9& 7.73& 113.7& 7.63\\
6496.460 &26.0& 4.795& -0.520& 0.0& 0.00& 93.8& 7.71& 95.6& 7.78& 95.4& 7.80& 0.0& 0.00\\
6498.935 &26.0& 0.960& -4.600& 125.1& 7.85& 0.0& 0.00& 120.5& 7.80& 0.0& 0.00& 0.0& 0.00\\
6518.363 &26.0& 2.830& -2.520& 0.0& 0.00& 0.0& 0.00& 99.3& 7.57& 108.6& 7.82& 110.9& 7.86\\
6533.925 &26.0& 4.560& -1.220& 75.5& 7.75& 0.0& 0.00& 0.0& 0.00& 0.0& 0.00& 0.0& 0.00\\
6556.783 &26.0& 4.795& -1.705& 0.0& 0.00& 33.3& 7.60& 37.0& 7.71& 36.5& 7.72& 0.0& 0.00\\
6569.212 &26.0& 4.733& -0.330& 0.0& 0.00& 109.5& 7.75& 110.4& 7.79& 108.7& 7.79& 105.5& 7.72\\
6574.222 &26.0& 0.990& -4.970& 94.9& 7.59& 92.6& 7.52& 96.1& 7.66& 90.4& 7.60& 89.3& 7.56\\
6581.203 &26.0& 1.480& -4.750& 0.0& 0.00& 82.1& 7.69& 86.4& 7.85& 0.0& 0.00& 0.0& 0.00\\
6591.304 &26.0& 4.590& -2.030& 0.0& 0.00& 0.0& 0.00& 31.9& 7.69& 0.0& 0.00& 32.6& 7.71\\
6593.870 &26.0& 2.430& -2.300& 139.9& 7.63& 139.7& 7.62& 137.6& 7.63& 140.1& 7.72& 143.2& 7.76\\
6608.020 &26.0& 2.280& -3.990& 61.8& 7.53& 67.1& 7.61& 59.9& 7.52& 61.1& 7.58& 55.0& 7.45\\
6625.014 &26.0& 1.010& -5.330& 0.0& 0.00& 0.0& 0.00& 0.0& 0.00& 0.0& 0.00& 88.3& 7.91\\
6627.537 &26.0& 4.550& -1.500& 60.4& 7.69& 58.0& 7.63& 58.4& 7.67& 59.3& 7.72& 65.7& 7.84\\
6633.406 &26.0& 4.835& -1.260& 0.0& 0.00& 0.0& 0.00& 0.0& 0.00& 61.9& 7.85& 56.6& 7.73\\
6633.743 &26.0& 4.560& -0.700& 0.0& 0.00& 102.4& 7.78& 98.3& 7.72& 91.9& 7.62& 97.6& 7.72\\
6699.132 &26.0& 4.590& -2.170& 22.6& 7.58& 0.0& 0.00& 27.0& 7.71& 19.1& 7.51& 0.0& 0.00\\
6703.563 &26.0& 2.760& -3.010& 90.5& 7.72& 92.3& 7.74& 85.9& 7.65& 81.0& 7.59& 78.8& 7.53\\
6713.736 &26.0& 4.790& -1.440& 45.0& 7.58& 0.0& 0.00& 43.5& 7.57& 40.8& 7.52& 0.0& 0.00\\
6725.352 &26.0& 4.100& -2.220& 42.8& 7.52& 45.2& 7.55& 40.8& 7.50& 46.8& 7.64& 44.7& 7.59\\
6726.662 &26.0& 4.610& -1.010& 72.5& 7.51& 73.9& 7.53& 68.6& 7.45& 72.9& 7.57& 66.8& 7.43\\
6733.145 &26.0& 4.640& -1.440& 55.1& 7.62& 0.0& 0.00& 54.0& 7.62& 58.5& 7.73& 47.3& 7.48\\
6739.516 &26.0& 1.557& -4.934& 62.7& 7.60& 58.9& 7.50& 57.1& 7.53& 0.0& 0.00& 58.6& 7.59\\
6745.951 &26.0& 4.070& -2.730& 0.0& 0.00& 0.0& 0.00& 25.1& 7.63& 0.0& 0.00& 0.0& 0.00\\
6750.149 &26.0& 2.420& -2.510& 133.1& 7.67& 128.0& 7.56& 126.4& 7.58& 124.4& 7.59& 127.4& 7.64\\
6752.700 &26.0& 4.638& -1.244& 0.0& 0.00& 71.6& 7.76& 0.0& 0.00& 74.2& 7.87& 0.0& 0.00\\
6786.853 &26.0& 4.190& -1.920& 57.3& 7.61& 58.6& 7.63& 54.9& 7.59& 54.6& 7.60& 52.0& 7.54\\
6793.251 &26.0& 4.070& -2.420& 44.1& 7.71& 0.0& 0.00& 37.8& 7.61& 35.4& 7.57& 0.0& 0.00\\
6806.839 &26.0& 2.730& -3.110& 87.5& 7.70& 87.8& 7.69& 89.1& 7.77& 81.3& 7.64& 82.9& 7.67\\
6810.256 &26.0& 4.610& -0.970& 80.5& 7.64& 70.4& 7.42& 74.5& 7.53& 77.6& 7.63& 68.4& 7.42\\
5197.570 &26.1& 3.230& -2.230& 102.2& 7.63& 102.9& 7.62& 104.9& 7.71& 0.0& 0.00& 108.1& 7.76\\
5234.623 &26.1& 3.221& -2.180& 97.2& 7.45& 103.6& 7.57& 103.3& 7.61& 102.7& 7.62& 0.0& 0.00\\
5264.800 &26.1& 3.230& -3.130& 0.0& 0.00& 60.0& 7.41& 61.7& 7.50& 0.0& 0.00& 0.0& 0.00\\
5325.550 &26.1& 3.221& -3.210& 0.0& 0.00& 54.2& 7.33& 63.3& 7.61& 65.5& 7.70& 65.1& 7.64\\
5425.245 &26.1& 3.199& -3.290& 68.3& 7.78& 65.8& 7.68& 61.9& 7.63& 70.3& 7.88& 67.2& 7.74\\
5991.366 &26.1& 3.150& -3.590& 53.8& 7.64& 59.0& 7.74& 57.0& 7.74& 50.3& 7.59& 44.5& 7.40\\
6084.096 &26.1& 3.200& -3.800& 38.5& 7.53& 53.0& 7.85& 36.7& 7.49& 0.0& 0.00& 42.1& 7.60\\
6149.239 &26.1& 3.890& -2.750& 47.9& 7.48& 0.0& 0.00& 50.5& 7.56& 49.5& 7.56& 46.8& 7.43\\
6247.557 &26.1& 3.870& -2.350& 62.0& 7.42& 66.6& 7.50& 73.0& 7.70& 64.4& 7.51& 73.7& 7.69\\
\hline
\end{tabular}
\end{table*}

\addtocounter{table}{-1} 
\begin{table*} 
\caption{Adopted line list, atomic parameters, equivalent widths (EW), and abundances (X)}
\begin{tabular}{ccrrrrrrrrrrrr}
\hline\hline
$\lambda$& Element& $\chi$& log(gf)& EW& X& EW& X& EW& X& EW& X& EW& X\\
\hline
&&&& 207 && 258 && 542 && 609 && 80 &\\
\hline
6369.453 &26.1& 2.890& -4.180& 0.0& 0.00& 32.7& 7.38& 40.3& 7.62& 36.7& 7.55& 0.0& 0.00\\
6432.675 &26.1& 2.890& -3.630& 66.1& 7.67& 57.4& 7.43& 61.9& 7.59& 54.2& 7.43& 59.1& 7.49\\
6456.379 &26.1& 3.900& -2.080& 80.8& 7.62& 79.3& 7.56& 81.1& 7.64& 77.5& 7.58& 78.3& 7.55\\
4935.825 &28.0& 3.941& -0.350& 73.6& 6.12& 65.7& 5.91& 70.3& 6.06& 64.9& 5.95& 75.8& 6.21\\
5010.930 &28.0& 3.635& -0.870& 76.8& 6.36& 79.3& 6.42& 74.9& 6.34& 74.8& 6.37& 0.0& 0.00\\
5094.403 &28.0& 3.830& -1.090& 54.5& 6.25& 55.1& 6.25& 53.3& 6.24& 52.1& 6.23& 42.9& 6.01\\
5102.960 &28.0& 1.676& -2.810& 95.6& 6.51& 95.7& 6.50& 94.4& 6.52& 87.6& 6.38& 0.0& 0.00\\
5155.120 &28.0& 3.898& -0.640& 78.4& 6.45& 69.7& 6.22& 78.8& 6.49& 0.0& 0.00& 0.0& 0.00\\
5197.155 &28.0& 3.898& -1.240& 59.6& 6.59& 0.0& 0.00& 70.4& 6.87& 0.0& 0.00& 0.0& 0.00\\
5468.100 &28.0& 3.850& -1.630& 36.3& 6.38& 0.0& 0.00& 36.4& 6.40& 31.3& 6.30& 0.0& 0.00\\
5578.716 &28.0& 1.676& -2.790& 0.0& 0.00& 0.0& 0.00& 0.0& 0.00& 0.0& 0.00& 89.0& 6.27\\
5589.353 &28.0& 3.898& -1.200& 54.1& 6.39& 49.5& 6.27& 55.5& 6.44& 0.0& 0.00& 53.4& 6.40\\
5748.347 &28.0& 1.680& -3.280& 78.3& 6.40& 71.5& 6.22& 74.5& 6.35& 74.4& 6.39& 83.6& 6.60\\
5760.826 &28.0& 4.100& -0.780& 65.1& 6.43& 64.7& 6.41& 65.8& 6.48& 61.8& 6.41& 56.5& 6.27\\
5996.724 &28.0& 4.230& -1.040& 0.0& 0.00& 46.4& 6.40& 46.3& 6.43& 42.9& 6.37& 40.5& 6.31\\
6007.306 &28.0& 1.676& -3.364& 74.6& 6.36& 72.4& 6.29& 66.7& 6.23& 65.9& 6.24& 64.1& 6.18\\
6086.274 &28.0& 4.260& -0.470& 69.6& 6.39& 71.0& 6.41& 73.7& 6.51& 69.9& 6.45& 69.9& 6.43\\
6108.112 &28.0& 1.680& -2.470& 116.3& 6.41& 114.3& 6.35& 111.5& 6.35& 113.0& 6.43& 114.9& 6.45\\
6111.067 &28.0& 4.090& -0.800& 65.1& 6.41& 55.6& 6.19& 61.2& 6.35& 58.5& 6.31& 60.6& 6.34\\
6128.972 &28.0& 1.680& -3.330& 84.4& 6.54& 78.9& 6.39& 80.4& 6.49& 79.9& 6.52& 78.3& 6.46\\
6130.129 &28.0& 4.260& -0.930& 51.9& 6.45& 47.1& 6.33& 48.0& 6.39& 40.9& 6.25& 38.4& 6.18\\
6176.806 &28.0& 4.090& -0.210& 92.5& 6.44& 92.3& 6.43& 89.3& 6.39& 97.3& 6.59& 92.5& 6.47\\
6177.240 &28.0& 1.830& -3.540& 53.0& 6.25& 54.9& 6.27& 53.3& 6.29& 49.8& 6.25& 57.6& 6.39\\
6204.599 &28.0& 4.090& -1.110& 47.3& 6.33& 48.6& 6.34& 53.5& 6.49& 47.8& 6.38& 42.9& 6.26\\
6223.979 &28.0& 4.100& -0.950& 55.0& 6.35& 49.7& 6.22& 53.6& 6.34& 52.9& 6.35& 57.4& 6.43\\
6322.161 &28.0& 4.150& -1.220& 39.8& 6.34& 41.4& 6.36& 45.7& 6.49& 39.9& 6.38& 37.3& 6.31\\
6327.596 &28.0& 1.680& -3.100& 99.9& 6.62& 97.2& 6.54& 93.0& 6.51& 92.3& 6.54& 98.9& 6.67\\
6378.246 &28.0& 4.150& -0.830& 63.2& 6.46& 52.9& 6.22& 64.1& 6.50& 60.4& 6.44& 60.3& 6.42\\
6482.795 &28.0& 1.930& -2.830& 101.9& 6.68& 100.9& 6.64& 107.8& 6.85& 98.1& 6.69& 100.0& 6.71\\
6586.306 &28.0& 1.950& -2.800& 89.5& 6.38& 90.3& 6.38& 96.3& 6.57& 92.8& 6.54& 93.0& 6.52\\
6598.594 &28.0& 4.230& -0.940& 48.2& 6.32& 50.0& 6.34& 52.0& 6.42& 51.0& 6.42& 48.9& 6.36\\
6635.117 &28.0& 4.420& -0.770& 49.7& 6.40& 52.8& 6.45& 54.0& 6.51& 51.9& 6.48& 56.8& 6.58\\
6767.769 &28.0& 1.830& -1.960& 128.5& 6.21& 131.3& 6.25& 130.6& 6.29& 125.8& 6.25& 133.3& 6.37\\
6772.311 &28.0& 3.660& -0.930& 83.6& 6.39& 89.1& 6.49& 86.1& 6.47&77.2& 6.31& 94.2& 6.65\\
\hline
\end{tabular}
\end{table*}

\end{document}